\documentclass[prd,preprintnumbers,nofootinbib,twocolumn,longbibliography]{revtex4-1}
\usepackage{amsmath}
\usepackage{graphicx}
\usepackage[bookmarksopen]{hyperref}

\makeatletter
\def\@bibdataout@aps{%
 \immediate\write\@bibdataout{%
  @CONTROL{%
   apsrev41Control%
   \longbibliography@sw{%
    ,author="08",editor="1",pages="1",title="0",year="0"%
   }{%
    ,author="08",editor="1",pages="0",title="",year="1"%
   }%
  }%
 }%
 \if@filesw
  \immediate\write\@auxout{\string\citation{apsrev41Control}}%
 \fi
}%
\makeatother

\newcommand\3[1]{\boldsymbol{#1}}

\newcommand\BIBvol[1]{\textbf{#1}}

\begin{document}
\title{The non-triviality of the vacuum in light-front quantization: An
       elementary treatment}
\author{John Collins}
\email{jcc8@psu.edu}
\affiliation{Department of Physics, Penn State University, University Park PA 16802, USA}
\date{11 January 2018}

\begin{abstract}
  It is often stated that the vacuum is trivial when light-front
  (null-plane) quantization is applied to a quantum field theory, in
  contrast to the situation with equal-time quantization.  In fact, it
  is has long been known that the statement is false, and that in
  certain cases the standard rules for light-front perturbation theory
  need modification.  This paper gives an elementary review of these
  issues, including an explanation of how and when there is a failure
  of the elementary derivation of the rules for light-front
  perturbation theory.
\end{abstract}

\maketitle

\section{Introduction}
\label{sec:intro}

It is commonly asserted (e.g., \cite{Brodsky:1997de, Heinzl:2000ht,
  Brodsky:2016nsn}) that the vacuum is trivial in a quantum field
theory constructed using light-front (null-plane) quantization, at
least within light-front perturbation theory.  This is unlike the
situation with standard Feynman perturbation theory.  Associated with
vacuum triviality appear to be a number of important consequences.
Among these is the possibility of making a useful and natural
definition of wave functions for particle states in the interacting
theory; the definition appears to arise directly from the property
that the state space of the theory is a free-particle Fock space.

Perhaps the most striking consequence that is claimed in
Ref.\ \cite{Brodsky:2009zd, Brodsky:2016nsn} is a solution of the
cosmological constant problem.  This solution arises because vacuum
energy bubbles appear to be zero in light-front perturbation theory
instead of being power-law divergent as they are in standard
calculations.  The vacuum bubbles give the vacuum expectation value of
the energy-momentum tensor, and hence a contribution to an effective
cosmological constant.  The divergence must be canceled by a
corresponding counterterm, a bare cosmological constant.  The
cosmological constant problem is that the value of the counterterm
must be extremely fine tuned.

However, it has been known for nearly 50 years, since the work of
Chang and Ma \cite{Chang:1968bh} and of Yan \cite{Yan:1973qg}, that
the argument leading to the triviality of the light-front vacuum is in
fact incorrect, as are the calculations of a zero value for vacuum
bubbles.  It was shown that the rules for light-front perturbation
theory must be modified, and that then the results always agree
between light-front and Feynman perturbation theory, including for
vacuum bubbles.  Further work, by Nakanishi and Yabuki
\cite{Nakanishi:1976yx}, and by Nakanishi and Yamawaki
\cite{Nakanishi:1976vf}, showed among other things that triviality of
the light-front vacuum conflicts with the well-established theorem of
Haag \cite{Haag:1955ev, Streater:1964}.  This theorem shows that the
representations of the commutation relations of field operators are
unitarily inequivalent between free and interacting
theories.\footnote{Technically, an extension of Haag's theorem is used
  \cite{Nakanishi:1976yx} so that it applies for the commutation
  relations on a null plane as well as for the equal-time commutation
  relations. 
}

In view of the continuing and prominent assertions of the triviality
of the light-front vacuum, the purpose of this paper is to give an
elementary treatment of the primary issues, especially concerning
actual calculations:
\begin{enumerate}

\item I review, using a very simple example, the demonstration that an
  inconsistency arises from the calculational method that gives
  vanishing of vacuum bubbles.  

\item I provide a new analysis to locate the failure in the derivation
  of the rules for light-front perturbation theory.  The failure is
  only for graphs (and subgraphs) for which the plus components of
  external momenta are constrained to be zero.

  The problem is an unrestricted use of the standard theorem $\int dx \,
  e^{ixq} = 2\pi \delta(q)$ that implements momentum conservation in terms of
  a delta function.  This formula is incorrect when integrated with a
  function that is discontinuous at $q=0$, as happens in the
  situations where there is a failure of the derivation of the
  standard rules for light-front perturbation theory.

\item I explain that, nevertheless, the non-triviality of the
  light-front vacuum does not itself affect the possibility of
  defining light-front wave functions.  (Other issues do intervene in
  a gauge theory or when a non-trivial ultra-violet field
  renormalization is needed.)

\end{enumerate}
This paper should be regarded as an elementary complement to Refs.\
\cite{Chang:1968bh, Yan:1973qg, Nakanishi:1976yx, Nakanishi:1976vf,
  Heinzl:2003jy, Herrmann:2015dqa}.  Some parts of the argument here
were already presented in a similar form in Sec.\ 7.2 of my QCD book
\cite{Collins:2011qcdbook}.  Most of the specific points made here are
undoubtedly known to many experts.  I hope the presentation here will
be useful to give an overall picture also accessible to interested
outsiders.

The discussion given here is in terms of the continuum theory.  The
issues appear in a different form when Discrete Light-Cone
Quantization is used \cite{Maskawa:1975ky, Tsujimaru:1997jt,
  Yamawaki:1998cy}.  These need a separate discussion that goes far
beyond the scope of the present paper.

\vfill 

\section{A paradox and its resolution in light-front perturbation theory}
\label{sec:paradox.resolve}

\subsection{Light-front perturbation theory; conventions}
\label{sec:conv}

We use light-front coordinates where components of a Lorentz vector
are defined as $x=(x^+,x^-,\3x_T)$, and where, given a choice of
Cartesian coordinates, $x^+=(x^0+x^3)/ \sqrt 2$, $x^-=(x^0-x^3)/ \sqrt
2$, and $\3x_T$ denotes the remaining transverse coordinates.

Graphs in light-front perturbation theory --- e.g.,
\cite{Brodsky:1997de} --- are specified with particular orderings in
the space-time coordinate $x^+$ for the vertices.  There are integrals
over the values of momentum components $k_j^+$ and $\3k_{j,T}$ for the
lines, constrained by conservation of the plus and transverse
components.  The values of $k^+$ are restricted to physical positive
values $k_j^+>0$ for forward moving momenta. The integrand has an
``energy denominator factor'' of the following form for each
intermediate state $I$
\begin{equation}
  \label{eq:denominator}
  \frac{i}{ P_{\rm ext}^- - \sum_{j\in I} \dfrac{k_{j,T}^2+m_j^2}{2k_j^+} + i\epsilon },
\end{equation}
where the sum is over particles in the intermediate state,
corresponding to lines in the graph.  Each denominator is the
difference between the external minus-component of momentum and the
on-shell plus-component of momentum for the intermediate state.  In
graphs, we use the convention that $x^+$ increases from left to right,
and correspondingly the flow of positive $k_j^+$ is from left to
right.  Note that plus momentum is conserved, so that $P_{\rm ext}^+ =
\sum_{j\in I} k_j^+$.

\begin{figure}
  \centering
    \includegraphics[scale=0.6]{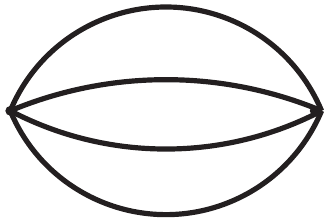}
  \caption{Example of vacuum bubble.}
  \label{fig:vac-bub}
\end{figure}

With the basic rules for light-front perturbation theory, vacuum
bubbles like Fig.\ \ref{fig:vac-bub} are zero.  This is because the
external momentum is zero.  Hence it has zero \emph{plus} component:
$P_{\rm ext}^+=0$, and there are no possible intermediate states: All
the lines have positive plus momentum from left to right, and these
can never sum to the zero external plus momentum.  The vanishing of
the vacuum bubbles is what leads to the statement of the triviality of
the vacuum.

\subsection{A paradox}
\label{sec:paradox}

In this section, I show how the basic rules for light-front
perturbation theory applied at vanishing external plus momentum lead
to an inconsistency with standard analyticity properties.  To make a
very simple situation, let us examine the connected part of the
following momentum-space Green function of composite operators in the
theory of a free scalar field of mass $m$:
\begin{equation}
  \label{eq:Pi}
  \Pi(p^2) = \int d^2x ~ e^{ip\cdot x}
            \langle 0 | T \tfrac12 \phi^2(x) \tfrac12 \phi^2(0) | 0 \rangle_{\rm conn.}.
\end{equation}
The factors of $1/2$ are to get a standard normalization convention,
and the use of the connected part means that from each instance of the
operator $\phi^2$ is subtracted its vacuum expectation value.  To make
the calculations maximally simple, while still exhibiting the
principles at stake, we work in $1+1$ space-time dimensions, without
any transverse dimensions.  A standard textbook property is that
$\Pi(p^2)$ is an analytic function whose only singularity is a branch
point at the threshold point $p^2=4m^2$.
If the second $\phi^2$ operator were at a general position $y$, we would
change $x$ in the exponent to $x-y$ without change in $\Pi(p^2)$, by
translation invariance.

The use of time-ordering of the operators in Eq.\ (\ref{eq:Pi}) might
appear to suggest the use of equal-time quantization.  However, it is
known that the definition is fully covariant.  When $x$ is
space-like, the operators commute, and then the ordering is
irrelevant.  When $x$ is time-like, it is frame-independent as to
which position is future-most, and then the corresponding operator is
defined to be on the left.  The situation with light-like separation
is governed by the usual rules for analyticity and for the
distributional nature of the fields.

A Green function of composite operators is used instead of an actual
S-matrix element to give maximum simplicity.  In QCD, we often use
such Green functions to formulate the strong-interaction part of a
scattering with electroweak particles, with the composite operators
corresponding to the coupling between an electroweak field and QCD
fields.

\begin{figure}
  \centering
  \addtolength{\tabcolsep}{6mm}
  \begin{tabular}{cc}
    
    \includegraphics[scale=0.6]{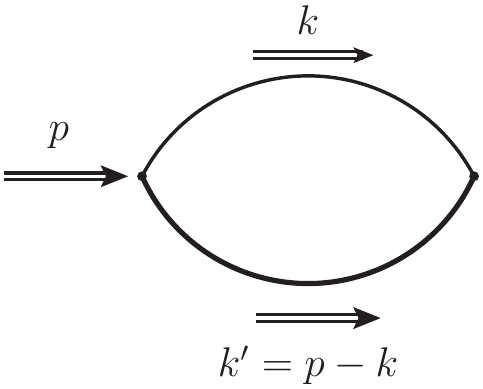}
  &
    \raisebox{3mm}{\includegraphics[scale=0.6]{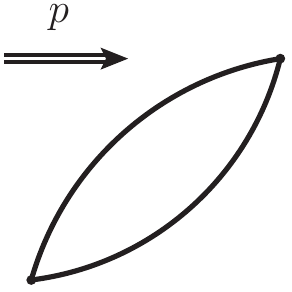}}
  \\
    (a) & (b)
  \end{tabular}
  \caption{The two orderings in $x^+$ for the Green function of two $\phi^2$ operators in
           free-field theory.
           Only graph (a) is non-zero when the external plus-momentum
           $p^+$ is positive.
         }
  \label{fig:loop}
\end{figure}

Given that free-field theory is used, there are exactly two
$x^+$-ordered graphs for $\Pi(p^2)$, as in Fig.\ \ref{fig:loop}.

When the external momentum has a positive plus component, i.e.,
$p^+>0$, the rules for $x^+$-ordered perturbation theory give a single
allowed ordering in $x^+$, symbolized in Fig.\ \ref{fig:loop}(a).  The
value of the graph, including its symmetry factor $1/2$, is
\begin{widetext}
\begin{align}
  \Pi(p^2) & = \frac{1}{4\pi}
      \int_0^{p^+} \frac{dk^+}{4k^+(p^+-k^+)} ~
      \frac{i}{ p^- - \frac{m^2}{2k^+} - \frac{m^2}{2(p^+-k^+)} + i\epsilon }
\nonumber\\
   & = \frac{i}{8\pi}
      \int_0^1 d\xi ~
      \frac{1}{ p^2 \xi(1-\xi) - m^2 + i\epsilon },
\label{eq:Pi.p+.gt.0}
\end{align}
\end{widetext}
where $\xi=k^+/p^+$.

When $p^+$ is negative, the other $x^+$ ordering, Fig.\
\ref{fig:loop}(b), is used and gives the same value for $\Pi(p^2)$.

But when $p^+=0$, the graphs appear to be zero, because the two
internal lines are required to have a physical momentum, with positive
plus momentum, and this is prohibited by momentum conservation.
However, we also know that the Green function is analytic at $p^2=0$,
and hence the value at $p^+=0$ is the limit as $p^2\to0$ of Eq.\
(\ref{eq:Pi.p+.gt.0}), which is $-i/(8\pi^2m^2)$ and non-zero.
Something is therefore wrong in the calculation at $p^+=0$.

Once we realize that in this example the standard rules for
light-front perturbation theory fail when the external momentum is
zero, we must then expect them to fail also in the computation of
vacuum bubbles.
Corresponding results also apply in a higher space-time dimensions
where there is also a transverse momentum integral.

\subsection{Diagnosis within Feynman perturbation theory}
\label{sec:diagnosis.F}

To diagnose the situation, consider the calculation of the same graph
in Feynman perturbation theory:
\begin{widetext}
\begin{align}
  \label{eq:Pi.Feyn}
  \Pi(p^2) &= -\frac{1}{8\pi^2} \int d^2k ~
      \frac{1}{ [k^2-m^2+i\epsilon] [(p-k)^2-m^2+i\epsilon] }
\nonumber\\
    &= -\frac{1}{8\pi^2} \int dk^+ dk^-
      \frac{1}{ [2k^+k^--m^2+i\epsilon] [2(k^+-p^+)(k^--p^-)-m^2+i\epsilon] }.
\end{align}
To get a result corresponding to light-front perturbation theory, we
perform the $k^-$ integral at fixed $k^+$.

We first take $p^+>0$, and use contour integration on $k^-$.  This
gives zero unless $0<k^+<p^+$, and then closing the contour in the
upper or lower half plane of $k^-$ gives the same result as was
calculated in light-front perturbation theory in Eq.\
(\ref{eq:Pi.p+.gt.0}).  This is an example of the established result
\cite{Chang:1968bh, Yan:1973qg} that in this case both methods of
calculation agree.

Next set $p^+=0$.  The integral is now
\begin{equation}
  \label{eq:Pi.Feyn.p+.0}
  \Pi(0)
    = -\frac{1}{8\pi^2} \int dk^+ dk^-
      \frac{1}{ [2k^+k^--m^2+i\epsilon] [2 k^+(k^--p^-)-m^2+i\epsilon] }.
\end{equation}
\end{widetext}
If $k^+$ is positive, then both $k^-$ poles are in the lower half
plane.  We can deform the contour to infinity in the opposite half
plane to get zero.  Similarly if $k^+$ is negative, both poles are in
the upper half plane, and we again get zero.

Hence it would appear that the graph is zero when $p^+=0$, in
agreement with the result of standard light-front perturbation theory.
But the result is surely wrong.  If nothing else, we could choose to
perform the $k^+$ integral first, which would correspond to obtaining
the result for the opposite kind of light-front perturbation theory,
i.e., ordered in $x^-$. Provided that $p^-$ is still non-zero, we get
the expected non-zero result for the graph, i.e., the limit of Eq.\
(\ref{eq:Pi.p+.gt.0}) as $p^2\to0$.

Evidently there is a mathematical error in evaluating the integral in
Eq.\ (\ref{eq:Pi.Feyn.p+.0}) by first performing the $k^-$ integral
and blindly using the zero result.  The correct result, as found by
Chang and Ma \cite{Chang:1968bh} and Yan \cite{Yan:1973qg}, is that
the integral over $k^-$ in Eq.\ (\ref{eq:Pi.Feyn.p+.0}) gives a
delta function at $k^+$.

\begin{figure}
  \centering
  \includegraphics[scale=0.55]{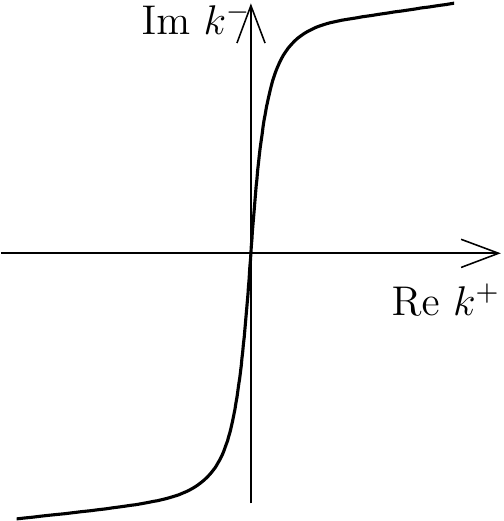}
  \caption{How to join the segments of contour used for positive and
    negative $k^+$ in evaluating Eq.\ (\ref{eq:Pi.Feyn.p+.0}). The
    horizontal axis is a slice at fixed $\text{Re} ~k^-$ from the
    2-dimensional space of $\text{Re} ~k^+$ and $\text{Re} ~k^-$ }
  \label{fig:contour.p+.0}
\end{figure}

To understand better what has happened, notice that we deformed the
$k^-$-contour infinitely far into the upper half plane when $k^+>0$
and infinitely far into the lower half plane when $k^+<0$.  The
two-dimensional contour over both of $k^+$ and $k^-$ is now broken at
$k^+=0$.  To complete the contour, we must couple these pieces by a
section that goes from $(k^+,k^-)=(0,+i\infty)$ to $(k^+,k^-)=(0,-i\infty)$, as
in Fig.\ \ref{fig:contour.p+.0}.  In deforming the unbroken
two-dimensional contour, we inevitably encounter a region where the
integrand is unsuppressed.


Another way of seeing the issue is to observe that when $k^+$ is fixed
and non-zero, and we deform $k^-$ to infinitely large imaginary $k^-$,
the integrand is of order $1/(k^-)^2$, and thus the contour at
infinity gives a zero integral.  But the coefficient of this asymptote
is $1/(k^+)^2$, and hence the convergence of the integrand to zero is
not uniform in $k^+$.  Instead we get an unsuppressed contribution if
we take $|k^-|$ to infinity while keeping $|k^+|$ of order $1/|k^-|$.
This problem does not arise in the integral for the case that $p^+$
non-zero, for then one of the $1/k^-$ factors has a coefficient
$1/(k^+-p^+)\to-1/p^+$ instead of $1/k^+$, and there is a suppression of
the otherwise dangerous region.

This dangerous region corresponds to modes with extremely large
negative rapidity which propagate almost within the surfaces of equal
$x^+$ on which light-front quantization is applied.  Thus they involve
large distances on these surfaces in the $x^-$ direction within the
surfaces.

Of course, the diagnosis just given relies on assuming that Feynman
perturbation theory is correct, while the authors of at least
\cite{Brodsky:1997de, Brodsky:2016nsn} give the impression that
light-front methods could give a different and better solution of a
theory.  If nothing else, they wish to formulate the theory purely in
terms of light-front methods and avoid any appeal to other methods.
Therefore to fully counter their arguments, one must understand the
situation solely within terms of light-front perturbation theory and
its derivation.

\section{What went wrong in derivation}
\label{sec:error}

In this section, I locate the error in the derivation of the basic
rules for light-front perturbation theory when the external $p^+$ is
zero, and I do this purely within light-front methods, avoiding
reference, for example, to standard Feynman perturbation theory.
Without correctly and precisely locating the error, there can be
generalized doubts as to the validity of light-front methods;
outsiders to the field can wonder what other errors are so far
unperceived.

\subsection{Momentum-conservation condition}

Textbook derivations of perturbation formalisms in quantum field
normally start with a coordinate space formulation and then use
Fourier transforms into momentum space.  This is what we do here.  In
formulating light-front perturbation theory, the field operators are
stratified by the values of their $x^+$ position coordinates.  Within
a surface of constant $x^+$ we perform the integral over $x^-$.  We
would also integrate over transverse coordinates if the problem were
in a higher space-time dimension than our example.

We apply this procedure to the graphs of Fig.\ \ref{fig:loop},
including both orderings in $x^+$.  Then, Eq.\ (\ref{eq:Pi.p+.gt.0})
is replaced by its effective predecessor in the derivation, which is
\begin{widetext}
\begin{multline}
\label{eq:vertex:x-}
  \Pi(p^2)  = \frac{1}{8\pi^2}
      \int_{-\infty}^\infty \frac{dk^+}{2k^+} 
      \int_{-\infty}^\infty \frac{d{k'}^+}{2{k'}^+} 
      \int_{-\infty}^\infty dx^- e^{ix^-(p^+-k^+-{k'}^+)}
      \left[
         \theta(k^+)\theta({k'}^+)
         \frac{i}{ p^- - \frac{m^2}{2k^+} - \frac{m^2}{2{k'}^+} + i\epsilon }
      \right.
\\
      \left.
         + \theta(-k^+)\theta({-k'}^+)
           \frac{i}{ - p^- + \frac{m^2}{2k^+} + \frac{m^2}{2{k'}^+} + i\epsilon }
       \right] .
\end{multline}
\end{widetext}
This equation applies for any value of $p^+$.  The integral over $x^+$
has already been performed to give the energy denominator factors.
The integrals over $k^+$ and ${k'}^+$ are over basis states
corresponding to the possible intermediate states. The $\theta$ functions
implement the physical-state conditions for the lines' momenta for
each of the $x^+$-ordered graphs.  The convention for the signs of
$k^+$ and ${k'}^+$ is chosen to give the same imaginary exponent for
both graphs.  Thus the positive, forward-moving momenta in Fig.\
\ref{fig:loop}(b) are $-k^+$ and $-{k'}^+$.

We now evaluate the integral over $x^-$.  Normally, we would obtain a
momentum-conservation delta function, from the theorem that
\begin{equation}
  \label{eq:delta}
  \int dx^- e^{ix^-(p^+-k^+-{k'}^+)} = 2\pi \delta(p^+-k^+-{k'}^+).
\end{equation}
This reproduces the previously given results for the graph in
light-front perturbation theory.  However, the theorem must be
interpreted in the sense of distributions, not as a theorem about
functions.  That is, the theorem applies provided that the left- and
right-hand sides are integrated with a function that is smooth enough,
In particular, the function must be continuous at $k^++{k'}^+=p^+$,
otherwise the result of integrating with the delta function is
undefined.

Now, if $p^+$ is non-zero, the rest of the integrand in Eq.\
(\ref{eq:vertex:x-}) is indeed continuous at the relevant points.  Then
Eq.\ (\ref{eq:delta}) applies, and the standard result is correct in
this case.

However when $p^+$ is zero, the integrand is not continuous on the
relevant line, which is $k^++{k'}^+=0$.  The integrand has
discontinuities when one or both of $k^+$ and ${k'}^+$ is zero, and
hence at the point $k^+={k'}^+=0$, on the line $k^++{k'}^+=0$.  In
fact, the integrand diverges as that point is approached from some
directions.  So we must investigate the situation in more detail.

We evaluate the integral by changing variables to $k^++{k'}^+$ and
$\xi$, with $k^+=\xi(k^++{k'}^+)$ and ${k'}^+=(1-\xi)(k^++{k'}^+)$.  The use
of $k^++{k'}^+$ ensures that one of the independent integration
variables is exactly the variable that appears in the exponential and
hence in the argument of the would-be delta function.  This gives
\begin{widetext}
\begin{equation}
\label{eq:vertex:x-.2}
  \Pi(p^2)  = \frac{1}{16\pi^2}
      \int_{-\infty}^\infty  d(k^++{k'}^+) \int_0^1 d\xi
      \int_{-\infty}^\infty dx^- e^{ix^-(p^+-k^+-{k'}^+)}
      \frac{i}{ 2p^-(k^++{k'}^+)\xi(1-\xi) - m^2 + i\epsilon }.
\end{equation}
\end{widetext}
The integrand is now smooth, even at $k^++{k'}^+=0$, and we can
therefore always use Eq.\ (\ref{eq:delta}) to give a delta function
that gives exactly the same value as the bottom line in Eq.\
(\ref{eq:Pi.p+.gt.0}).  But now the derivation is valid for any $p^+$,
including zero, and this was arranged by using a coordinate-space
expression as a starting point.

\subsection{Immediate implications}

When $p^+$ is zero, the integral over $x^-$ in Eq.\
(\ref{eq:vertex:x-.2}) gives a delta function at $k^++{k'}^+=0$. Then
because $k^+$ and ${k'}^+$ are restricted to have the same sign, and
because $\xi$ is bounded, the values of $k^+$ and ${k'}^+$ that are
relevant are both zero.  Thus we have a kind of zero-mode
contribution.

When one or both of $k^+$ and ${k'}^+$ goes to zero, the energy
denominators in (\ref{eq:vertex:x-}) go to infinity.  This would give
a zero in the integrand were it not for a divergence in the factor
$1/(4k^+{k'}^+)$ that is associated with the Lorentz invariant
integration over the momenta of the lines.  The combined limiting
behavior from these factors and the Jacobian of the transformation of
variables gives the non-zero final result.  The distributional nature
of the calculation with an integral over all plus-momenta as a
starting point allows the calculation to involve a limiting behavior
from nonzero values of plus-momenta.

Zero plus-momentum implies infinitely large minus-momentum, and hence
infinitely large and negative rapidity for each of the lines.  Thus
the physical contribution concerns contributions from intermediate
states with arbitrarily negative rapidities.

Smoothness (and analyticity, in fact) of the integrand in
(\ref{eq:vertex:x-.2}) is only obtained after summing both $x^+$
ordered graphs in Fig.\ \ref{fig:loop}.  This results from the
relationship between the two energy denominators in
(\ref{eq:vertex:x-}), which is presumably a fundamental property
related to the CPT invariance of relativistic quantum field theories.
But I will leave that issue to others to investigate in generality.

There is a difference in the definition of $\xi$ compared with the one
used in Eq.\ (\ref{eq:Pi.p+.gt.0}).  There $\xi$ was defined as
$k^+/p^+$, which does not work when $p^+$ is zero.  Here it is defined
as $k^+/(k^++{k'}^+)$, which does not depend on $p^+$.  Of course,
after application of the delta function the definitions agree, but
only when $p^+$ is strictly positive.

\subsection{General case}

The calculation just given is, of course, specific to one particular
pair of graphs.  But we can now draw some more general conclusions.

First, the problem does not arise if we use time-ordered instead of
$x^+$-ordered perturbation theory.  This because it is only
$x^+$-ordered perturbation theory that has boundaries on the allowed
range of physical momenta.  

As for a general graph in the $x^+$-ordered case, each vertex has a
factor that is a simple generalization of the $x^-$ integral in Eq.\
(\ref{eq:vertex:x-}).  To avoid a long technical discussion, let us
leave to other work a general derivation of the final answer strictly
using only light-front methods.  Here we just appeal to Refs.\
\cite{Chang:1968bh, Yan:1973qg}.  There it was shown by derivation
from Feynman graphs that the only cases when the basic rules fail to
be valid is in graphs where the momentum-conservation conditions
require that one or more lines are constrained to have vanishing
plus-momenta, just as in our example graphs at $p^+=0$ and in vacuum
bubbles.  Those references showed how to obtain corrected rules for
light-front perturbation theory.

\subsection{Overall summary}

We can now summarize the results of this section
\begin{quote}
  The failure of the standard derivation of the basic rules for
  light-front perturbation theory occurs where there is a failure of
  theorems like Eq.\ (\ref{eq:delta}) that give delta-functions for
  momentum conservation.  Momentum conservation does continue to hold.
  But when some lines are constrained to be at zero plus-momentum, the
  standard delta function must be changed to a different kind of delta
  function that correctly takes into account the limiting behavior as
  the zero mode configuration is approached.
\end{quote}

\section{Consequences, or lack thereof, for light-front wave
  functions}
\label{sec:lfwf}

A conspicuous contrast between relativistic quantum field theory and
non-relativistic quantum mechanics concerns the status and use of wave
functions.  Their use is routine in atomic, molecular, and condensed
matter physics, and gives a lot of useful information about the
states.  For example, crucial insights into new structures emerging
for interacting electrons in condensed matter physics has arisen from
thinking about wave functions, two prominent examples being
superconductivity \cite{DeGennes.1966} and the fractional quantum Hall
effect \cite{Jain2007.book}.

In contrast, in work with relativistic quantum field theory
there is little systematic use of wave functions, and correspondingly
a paucity of information about the detailed microscopic nature of the
quantum mechanical states involved.  This is a particular important
issue for non-perturbative bound states in QCD, i.e., for hadrons and
nuclei.

The one known exception is with light-front quantization.  Interesting
progress has been made in applications, e.g., \cite{Brodsky:2016nsn,
  Alberg:2017ijg, Hiller:2016itl, Vary:2016ccz, Vary:2016emi} and references therein.
However, the standard formulation, as in \cite{Brodsky:1997de},
appears to rely on the vacuum being trivial, so that a given state can
be expressed as a non-pathological sum and integral over basis states
that are in the same Fock space as free particles, e.g., for a proton
in QCD in terms of states of quarks and gluons.

In general, a wave function gives an expression for a
quantum-mechanical state in terms of a set of basis states, which are
fixed independently of the presence and nature of the interactions in
a theory.  Thus the basis is defined before one has found a solution
of the theory.  In a non-relativistic system the basis is commonly of
eigenstates of the position operators, but a momentum-space basis
could also be used.  It gives a representation of the canonical
relations for the fundamental operators, which are coordinates and
canonical momenta in a standard non-relativistic system.

But in a relativistic theory, these natural ideas appear to conflict
with Haag's theorem.  Inequivalent representations of the canonical
commutation relations are needed in the free and interacting theories.
In a sense, the states for the interacting theory are in a different
space than those of a corresponding free theory.  One appears to need
to solve the theory before knowing what the basis states are;
equivalently, finding a solution of the theory includes a construction
of its state space.

In this section, I show how light-front quantization evades this issue
sufficiently to allow a definition of wave functions even given
non-triviality of the vacuum.  In a sense the difficulties are all
confined to the nature of the vacuum.  Haag's theorem itself cannot be
prevented from applying, but its consequences can be limited.

\subsection{Overall framework}
\label{sec:wf.overall}

To be able to easily relate different viewpoints, let us use the
Heisenberg picture.  We conceive of a theory being defined in terms of
operators and commutation\footnote{Anticommutation relations in the
  case of fermionic fields.} relations specified on a quantization
surface (e.g., fixed time or fixed $x^+$), and then evolution in $t$
or $x^+$ is applied.  In the Heisenberg picture, it is the operators
that are evolved rather than the states.

We now work in a field theory framework.  Suppose that we are able to
define annihilation and creation operators $a_{\3k}$ and $a^\dagger_{\3k}$
from the Fourier transform of the fields on the quantization surface,
that they obey the standard commutation relations for annihilation and
creation operators\footnote{Derived from the canonical commutation
  relations of the fields on the quantization surface.}, and that the
annihilation operators give zero when acting on the vacuum:
\begin{equation}
  \label{eq:wf.vac.annih}
  a_{\3k} |0\rangle = 0.
\end{equation}
For the purposes of this discussion we assume there is only one kind
of each operator; the extension to multiple fields is elementary.

The most general state constructed by applying creation operators to
the vacuum, has the form
\begin{equation}
  \label{eq:wf.expansion}
  |\psi\rangle = \sum_N \frac{1}{N!} \int \prod_{j=1}^N a^\dagger_{\3k_j} |0\rangle \psi_{\3k_1,\ldots,\3k_N} ,
\end{equation}
with an integration measure appropriate to the momentum variables and
the normalization of the operators.  The numerical coefficient
functions we call the momentum-space wave functions of the state.
Coordinate-space wave functions are defined by Fourier transformation
of the momentum-space wave functions.

Then the wave functions of a state can be obtained from a matrix
element of annihilation operators between the vacuum and the state:
\begin{equation}
  \label{eq:wf.basic}
  \psi_{\3k_1,\ldots,\3k_N} = \langle0| \prod_{j=1}^N a_{\3k_j} |\psi\rangle.
\end{equation}
This is proved using the commutation relations for the annihilation
and creation operators, together with the vacuum-annihilation
condition.

In the case of non-relativistic systems, wave functions with the above
definition are the same as ordinary Schr\"odinger wave functions.  This
is shown by the equivalence, e.g., \cite{Brown:1992db}, between a
theory of non-relativistic Schr\"odinger field(s) and a collection of
Schr\"odinger wave function theories for arbitrarily many particles.

Given the expression for the annihilation operators in terms of the
fields, the right-hand side of Eq.\ (\ref{eq:wf.basic}) can be
expressed as a Fourier transform of a corresponding matrix element of
field operators restricted to the quantization surface.  Thus the
formula can be applied independently of the method by which
calculations are made. For example, with a surface of fixed $x^+$ one
can equivalently use Feynman perturbation theory or $x^+$-ordered
perturbation theory.\footnote{But note that there are some further
  complications that need to be handled in gauge theories --- Sec.\
  \ref{sec:wf.complications}.}

Formulas like the above, or some equivalent, can be found in many
places in the literature, e.g., in the review \cite{Brodsky:1997de}
for the case of light-front quantization. Our ability to use them now
depends on whether or not we can find a useful definition of the
annihilation and creation operators in terms of the field operators.

As regards the implications of Haag's theorem, we now see an
interesting change of status of a wave function between
non-relativistic and relativistic theories.  In a non-relativistic
theory, the state space is the same with and without interactions, and
the vacuum is the same state.  In a relativistic theory, provided that
the vacuum-annihilation condition is obeyed, we have a \emph{labeling}
of a standard set of basis states that is the same as in the free
theory, as shown by Eq.\ (\ref{eq:wf.expansion}).  To avoid a conflict
with Haag's theorem, one could have the free and interacting vacua
being different.  Alternatively, as a mathematical device one could
take the abstract state spaces to be the same in the free and
interacting theories; but then some pathologies must arise in how the
free and interacting fields act on this space; that is, in how one
constructs fields obeying the equations of motion in the free and
interacting theories.  See \cite{Schlieder:1972qr, Herrmann:2015dqa}
for explanations of how this works out in terms of the field operators
when light-front quantization is used: essentially the complications
are avoided in light-front quantization by projecting out of the field
operator its behavior at $k^+=0$.

For the case of fields specified on a surface of fixed time $t$, there
is no way of defining annihilation and creation operators by Fourier
transformation of fields on the surface, such that they give the
vacuum-annihilation conditions.  Certainly none has been discovered;
see the literature on light-front quantization for details.

\subsection{Wave functions and light-front quantization}
\label{sec:wf.lf}

Compared with equal-time quantization, the situation radically changes
in light-front quantization.  On a surface of fixed $x^+$, we define
annihilation and creation operators by Fourier transformation:
\begin{widetext}
\begin{align}
\label{eq:scalar.lf}
  \phi(x^+,x^-,\3{x}_T) = {}& \int \frac{dk^+d^2\3k_T}{16\pi^3|k^+|}
          \theta(k^+)
          \left( a_{k^+,\3k_T}(x^+) e^{-ik^+x^- + i \3k_T\cdot\3x_T}
                 + a^\dagger_{k^+,\3k_T}(x^+) e^{ik^+x^- - i \3k_T\cdot\3x_T}
          \right)
\nonumber\\
  = {}& \int \frac{dk^+d^2\3k_T}{16\pi^3|k^+|}
          e^{-ik^+x^- + i \3k_T\cdot\3x_T}
          \left( a_{k^+,\3k_T}(x^+) \theta(k^+) + a^\dagger_{-k^+,-\3k_T}(x^+) \theta(-k^+)
          \right).
\end{align}
\end{widetext}
The key thing is that values of physical plus momenta are restricted
to be positive, and we express this by the allowed values for $k^+$ in
each term. In the second line, it is arranged to have a common
exponential factor, and there we can distinguish annihilation and
creation operators by the sign of $k^+$.  (Observe the reversal of
sign of the argument of $a^\dagger$ between the two lines.)  The explicit
denominator factor of $|k^+|$ gives a Lorentz-invariant form of the
integral over momenta.  The annihilation and creation operators are
given dependence on $x^+$, which is determined by the solution of the
theory.

It follows from Eq.\ (\ref{eq:scalar.lf}) that annihilation and
creation operators can be obtained from the field operator by a
Fourier transformation on a surface of fixed $x^+$, given a non-zero
value of $k^+$.  The standard commutation relations for the
annihilation and creation operators follow from the commutation
relation of the field on a surface of fixed $x^+$.  Furthermore,
applying an annihilation operator (with nonzero $k^+$) to the vacuum
gives zero.  This is simply because by the known properties of the
field under translations (in the full theory including interactions),
the state $a_{k^+,\3k_T}|0\rangle$ would have negative plus momentum
relative to the vacuum, which is not possible.

Thus all the conditions summarized in Sec.\ \ref{sec:wf.overall} for
defining wave functions are obeyed, and we have not had to invoke
triviality of the vacuum to do this.  

An immediate complication is that in a field theory, the most general
state is obtained by repeatedly applying field operators to the field
and taking linear combinations.  Normally it would be sufficient to
restrict the field operators to the quantization surface, and this
would show that the most general state is of the form
(\ref{eq:wf.expansion}).  The field operators get integrated with
coordinate-space wave functions.  This is compatible with the property
that the fields are not actual operators but are operator-valued
distributions; operators are only defined with an integration with a
smooth test function. But in light-front quantization, the creation
(and annihilation) operators are restricted to non-zero $k^+$.  When
the algebra of the operators is examined \cite{Schlieder:1972qr,
  Herrmann:2015dqa}, this corresponds to a restriction to using test
functions whose Fourier transform is zero at zero $k^+$.
Correspondingly, the wave functions $\psi(k_1^+,\3k_{1,T};
k_2^+,\3k_{2,T}; \ldots)$ have to vanish when one or more $k_j^+$ is zero.

There is the potential for extra states obtained by applying zero-mode
operators on the vacuum.  To implement this properly in terms of field
operators one must go slightly off the light-front
\cite{Schlieder:1972qr, Herrmann:2015dqa}.  In momentum-space, the use
of the appropriate distribution-theoretic framework indicates that the
implementation of the extra zero-mode contributions uses integrals
over a neighborhood of $k^+=0$, rather like that found in our earlier
calculation in Eq.\ (\ref{eq:vertex:x-.2}) at $p^+=0$. These are high
rapidity modes, and presumably susceptible to a systematic analysis
like those used in factorization or Regge theory.  Further examination
is needed.

\subsection{Complications}
\label{sec:wf.complications}

Beyond general zero-mode issues of the kind just discussed, there are
two complications that require modifications of the framework. One is
to deal with non-trivial field renormalization, and the other is for
gauge theories.  Neither of these applies in a super-renormalizable
non-gauge model, e.g., Yukawa theory in $2+1$ space-time dimensions.

\subsubsection{Renormalization}

The commutation relations that give the standard interpretation of
annihilation and creation operators are derived from the commutation
relations of the fields on the quantization surface. But in a
renormalizable theory, the fields need a renormalization factor:
\begin{equation}
\label{eq:field.ren}
    \phi_0(x) = \sqrt{Z} \phi_R(x).
\end{equation}
Here $\phi_0$ is the bare field, having the standard normalization for
its commutation relations, and $\phi_R$ is the renormalized field.  It is
the renormalized field whose Green functions and matrix elements are
finite.  To define the theory an ultra-violet cut-off is applied.
Order by order in perturbation theory, the renormalization factor $Z$
diverges.  The divergences can be resummed by renormalization-group
methods, at least if the theory is asymptotically free.  

When an on-shell ``physical'' renormalization prescription is used to
define the normalization of $Z$, the K\"allen-Lehmann representation for
the two-point functions can be used to show that the exact value of
$Z$ obeys $0\leq Z<1$, a fact reflected in the sign of its anomalous
dimension, and in a negative value for the lowest-order correction to
$Z$ in perturbation theory.  In the case that the lowest-order
divergence is at one-loop order, the renormalization group shows that
the exact value obeys $Z\to0$ as the cut-off is removed, in an
asymptotically free theory.

To get finite wave functions, we must apply the same renormalization
factor to the annihilation and creation operators, with the outcome
that the renormalized annihilation operators used to calculate
finite wave functions in Eq.\ (\ref{eq:wf.expansion}) are each an
infinite factor $1/\sqrt{Z}$ times the ones obey the standard
commutation relations.  This needs a modification of the formula
(\ref{eq:wf.expansion}) for a state.  I am not aware of a systematic
treatment of this issue.  

It is possible to do all calculations with a cutoff that is not
removed.  But it is surely preferable to say that the theory itself is
the renormalized theory defined in the limit that the cutoff is
removed.  Then Eq.\ (\ref{eq:wf.basic}) gives a valid definition of
renormalized wave functions in terms of matrix elements of
renormalized operators.  In situations where perturbation theory
applies, perturbative calculations of wave functions work.  But the
actual expansion of the quantum mechanical state needs modification
from (\ref{eq:wf.expansion}).  This is presumably a purely technical
problem, since ultra-violet renormalization is very well understood.

\subsubsection{Rapidity divergences in gauge theory}

Much more interesting is the problem with rapidity divergences in any
gauge theory.  

The natural gauge condition to use with quantization on a plane of
constant $x^+$ is \cite{Kogut:1969xa} the light-cone gauge $A^+=0$,
which gives the simplest version of the formalism.  Unfortunately, the
wave functions defined using this method have divergences
\cite{Ma:2006dp} beyond those associated with ultra-violet
renormalization.  These divergences have exactly the same cause as
those that arise when the same annihilation and creation operators are
used in the natural way to try to define transverse-momentum-dependent
(TMD) parton densities.  The divergences can be seen readily in
perturbative calculations, when going beyond lowest order.  The
divergences arise from regions of integration where rapidities of the
lines for the gauge fields go to negative infinity.

For example, Brodsky et al.\ \cite{Brodsky:2000ii} made calculations
in QED of the light-front wave functions of a single electron.  The
one-electron component is given in their Eq.\ (25) in terms of a
quantity they call $Z$ (not to be confused with the renormalization
factor in Eq.\ (\ref{eq:field.ren}) in the present paper).  The
one-loop value is given in their Eq.\ (29).  It has an integral over a
variable $x$ from 0 to 1, and the integral diverges logarithmically at
$x=1$. This contrasts with the statement just after (25) that $Z$ is
finite when the theory is regulated in the ultraviolet and infrared.

That such rapidity divergences are a general phenomena can be seen by
the same methods that Collins and Soper \cite{Collins:1981uk,
  Collins:1981uw} devised for the TMD parton densities and
fragmentation functions.  To regulate the divergences they changed to
a definition \cite{Soper:1979fq} that uses the same formula in terms
of field operators but with the use of space-like axial gauge $n\cdot A=0$.
The matrix elements depend on an extra parameter $\zeta$.  Collins and
Soper derived an equation for the $n$-dependence and hence the
$\zeta$-dependence that is phenomenologically important.  When $n$ becomes
light-like, i.e., when $n^\mu\to(0,1,\30_T)=g^{\mu+}$, the parameter $\zeta$
goes to infinity.  The Collins-Soper equation shows in full generality
that there is a divergence in this limit.

An alternative method gives an explicitly gauge invariant definition.
It starts by writing operator matrix elements such as the one in
(\ref{eq:wf.basic}) in terms of integrals over gauge-invariant field
operators.  In the expression in terms of operators in the light-cone
gauge, one inserts Wilson lines in the minus direction, to make the
operators gauge invariant. In the $A^+=0$ gauge, the Wilson lines are
unity, but we now have a definition that can be used with any gauge
condition.  Calculations reproduce the same rapidity divergences found
earlier.  A new method \cite{Collins:2011qcdbook} was devised to give
a kind of rapidity renormalization factor using vacuum matrix elements
of certain specially chosen Wilson loops.  The resulting matrix
elements have a parameter $\zeta$ with the same significance Collins and
Soper's $\zeta$, and the evolution equations are of the same form.  For
related definitions in soft-collinear effective theory (SCET), see
\cite{Becher:2010tm, GarciaEchevarria:2011rb, Collins:2012uy}.

Although these methods were devised in the context of TMD parton
densities, the principles apply equally to light-front wave functions,
as was shown by Ma and Wang \cite{Ma:2006dp}.  An appropriate version
of the modern definitions was given by Li and Wang \cite{Li:2014xda}.

Note the these considerations apply to the kind of light-front wave
function that has dependence on transverse momenta.  In contrast, much
phenomenology is done with a different kind of distribution amplitude
that is integrated over transverse momentum.  For these, rapidity
divergences cancel, and the evolution equation is a kind of
renormalization-group equation, the Efremov-Radyushkin-Brodsky-Lepage
(ERBL) equation.  This is the case at least when Radyushkin's
definition \cite{Radyushkin:1977gp,Radyushkin:2009zg} is used.
However, the Brodsky-Lepage definition \cite{Lepage:1979zb,
  Lepage:1980fj} is made in light-cone gauge with an explicit cutoff
in transverse-momentum.  If that definition is taken literally and its
implications examined closely enough, it is expected to give rapidity
divergences.

\section{Discussion}
\label{sec:discuss}

The use of light-front quantization has a number of important
advantages \cite{Chang:1968bh, Brodsky:1997de, Heinzl:2000ht},
compared with equal-time quantization. Many are presented
\cite{Brodsky:2016nsn} as being directly related to the vacuum being
trivial.  An interesting consequence of vacuum triviality is that the
effective cosmological constant caused by vacuum bubbles is zero
\cite{Brodsky:2009zd, Brodsky:2016nsn}, thereby trivially solving the
notorious cosmological constant problem.  But this solution comes at
the price that it implies an inequivalence between the solution of a
quantum field theory by light-front quantization and by conventional
methods.

However, it has long been known that the claim of vacuum triviality is
wrong \cite{Chang:1968bh, Yan:1973qg, Nakanishi:1976yx,
  Nakanishi:1976vf}. Furthermore the validity of the inconsistency
between the results of different kinds of quantization was challenged
by the demonstration \cite{Chang:1968bh, Yan:1973qg} that an actual
paradox is produced by the method of calculation that gives the
vanishing vacuum bubbles.  Chang and Ma \cite{Chang:1968bh} and Yan
\cite{Yan:1973qg} derived corrected rules for light-front perturbation
theory.  The corrections are confined to situations typified by the
vacuum bubbles used to calculate the bare contribution to an effective
cosmological constant.  But their starting point was an assumption
that Feynman perturbation theory is correct.  In view of a possible
inequivalence between different methods of quantization, this is not a
sufficient argument.

Therefore, motivated by continuing assertions about vacuum triviality
and the cosmological constant, the present paper tried to give an
elementary account that I hope will lay this issue to rest.  First, I
gave an example of the paradox mentioned above, and explained why it
indicates that there is almost certainly an error in the rules that
lead to the vanishing of vacuum bubbles, etc.  Then I located the
error in the derivation of the rules for light-front perturbation
theory.  The flaw is in the unrestricted use of a standard theorem to
get a delta-function to implement momentum conservation.  In exactly
the conditions needed for the cosmological constant calculation, the
theorem needs to be changed. The change restores the equivalence
between the results in different methods, in accordance with the old
results of Chang and Ma and Yan, but without needing the starting
assumption that it is Feynman perturbation theory that is correct.
Equally, the results now support non-triviality of the vacuum.

As we saw in Sec.\ \ref{sec:lfwf}, non-triviality of the vacuum does
not affect the ability to define light-front wave functions.  However,
in a gauge theory, the standard definitions have rapidity divergences,
and modified definitions are compulsory to deal with this
\cite{Ma:2006dp, Li:2014xda}, with complete similarity to
corresponding issues in the definition and use of
transverse-momentum-dependent parton densities.  The divergences to be
dealt with in this fashion arise from an integral over the rapidity of
gluonic configurations, and give divergences when the plus momenta of
these configurations go to zero.


\begin{acknowledgments}
  This work was supported in part by the U.S. Department of Energy
  under Grant No.\ DE-SC0013699.  I would like to thank Stan Brodsky,
  Thomas Heinzl, Paul Hoyer, Jainendra Jain, and Ted Rogers for useful
  discussions.
\end{acknowledgments}


\addtolength{\textheight}{3mm} 

\bibliography{jcc}

\providecommand{\noopsort}[1]{}
\begin{thebibliography}{38}%
\makeatletter
\providecommand \@ifxundefined [1]{%
 \@ifx{#1\undefined}
}%
\providecommand \@ifnum [1]{%
 \ifnum #1\expandafter \@firstoftwo
 \else \expandafter \@secondoftwo
 \fi
}%
\providecommand \@ifx [1]{%
 \ifx #1\expandafter \@firstoftwo
 \else \expandafter \@secondoftwo
 \fi
}%
\providecommand \natexlab [1]{#1}%
\providecommand \enquote  [1]{``#1''}%
\providecommand \bibnamefont  [1]{#1}%
\providecommand \bibfnamefont [1]{#1}%
\providecommand \citenamefont [1]{#1}%
\providecommand \href@noop [0]{\@secondoftwo}%
\providecommand \href [0]{\begingroup \@sanitize@url \@href}%
\providecommand \@href[1]{\@@startlink{#1}\@@href}%
\providecommand \@@href[1]{\endgroup#1\@@endlink}%
\providecommand \@sanitize@url [0]{\catcode `\\12\catcode `\$12\catcode
  `\&12\catcode `\#12\catcode `\^12\catcode `\_12\catcode `\%12\relax}%
\providecommand \@@startlink[1]{}%
\providecommand \@@endlink[0]{}%
\providecommand \url  [0]{\begingroup\@sanitize@url \@url }%
\providecommand \@url [1]{\endgroup\@href {#1}{\urlprefix }}%
\providecommand \urlprefix  [0]{URL }%
\providecommand \Eprint [0]{\href }%
\providecommand \doibase [0]{http://dx.doi.org/}%
\providecommand \selectlanguage [0]{\@gobble}%
\providecommand \bibinfo  [0]{\@secondoftwo}%
\providecommand \bibfield  [0]{\@secondoftwo}%
\providecommand \translation [1]{[#1]}%
\providecommand \BibitemOpen [0]{}%
\providecommand \bibitemStop [0]{}%
\providecommand \bibitemNoStop [0]{.\EOS\space}%
\providecommand \EOS [0]{\spacefactor3000\relax}%
\providecommand \BibitemShut  [1]{\csname bibitem#1\endcsname}%
\let\auto@bib@innerbib\@empty
\bibitem [{\citenamefont {Brodsky}\ \emph {et~al.}(1998)\citenamefont
  {Brodsky}, \citenamefont {Pauli},\ and\ \citenamefont
  {Pinsky}}]{Brodsky:1997de}%
  \BibitemOpen
  \bibfield  {author} {\bibinfo {author} {\bibfnamefont {S.~J.}\ \bibnamefont
  {Brodsky}}, \bibinfo {author} {\bibfnamefont {H.-C.}\ \bibnamefont {Pauli}},
  \ and\ \bibinfo {author} {\bibfnamefont {S.~S.}\ \bibnamefont {Pinsky}},\
  }\bibfield  {title} {\enquote {\bibinfo {title} {Quantum chromodynamics and
  other field theories on the light cone},}\ }\href@noop {} {\bibfield
  {journal} {\bibinfo  {journal} {Phys. Rept.}\ }\textbf {\bibinfo {volume}
  {301}},\ \bibinfo {pages} {299--486} (\bibinfo {year} {1998})},\ \Eprint
  {http://arxiv.org/abs/hep-ph/9705477} {hep-ph/9705477} \BibitemShut {NoStop}%
\bibitem [{\citenamefont {Heinzl}(2001)}]{Heinzl:2000ht}%
  \BibitemOpen
  \bibfield  {author} {\bibinfo {author} {\bibfnamefont {T.}~\bibnamefont
  {Heinzl}},\ }\bibfield  {title} {\enquote {\bibinfo {title} {Light-cone
  quantization: {F}oundations and applications},}\ }\href@noop {} {\bibfield
  {journal} {\bibinfo  {journal} {Lect. Notes Phys.}\ }\textbf {\bibinfo
  {volume} {572}},\ \bibinfo {pages} {55--142} (\bibinfo {year} {2001})},\
  \Eprint {http://arxiv.org/abs/hep-th/0008096} {hep-th/0008096} \BibitemShut
  {NoStop}%
\bibitem [{\citenamefont {Brodsky}(2016)}]{Brodsky:2016nsn}%
  \BibitemOpen
  \bibfield  {author} {\bibinfo {author} {\bibfnamefont {S.~J.}\ \bibnamefont
  {Brodsky}},\ }\bibfield  {title} {\enquote {\bibinfo {title} {Light-front
  holography, color confinement, and supersymmetric features of {QCD}},}\
  }\bibfield  {booktitle} {\emph {\bibinfo {booktitle} {{Proceedings, Theory
  and Experiment for Hadrons on the Light-Front (Light Cone 2015): Frascati ,
  Italy, September 21-25, 2015}}},\ }\href {\doibase 10.1007/s00601-016-1070-8}
  {\bibfield  {journal} {\bibinfo  {journal} {Few Body Syst.}\ }\textbf
  {\bibinfo {volume} {57}},\ \bibinfo {pages} {703--715} (\bibinfo {year}
  {2016})},\ \Eprint {http://arxiv.org/abs/1601.06328} {arXiv:1601.06328
  [hep-ph]} \BibitemShut {NoStop}%
\bibitem [{\citenamefont {Brodsky}\ and\ \citenamefont
  {Shrock}(2011)}]{Brodsky:2009zd}%
  \BibitemOpen
  \bibfield  {author} {\bibinfo {author} {\bibfnamefont {S.~J.}\ \bibnamefont
  {Brodsky}}\ and\ \bibinfo {author} {\bibfnamefont {R.}~\bibnamefont
  {Shrock}},\ }\bibfield  {title} {\enquote {\bibinfo {title} {{Condensates in
  Quantum Chromodynamics and the Cosmological Constant}},}\ }\href {\doibase
  10.1073/pnas.1010113107} {\bibfield  {journal} {\bibinfo  {journal} {Proc.
  Nat. Acad. Sci.}\ }\textbf {\bibinfo {volume} {108}},\ \bibinfo {pages}
  {45--50} (\bibinfo {year} {2011})},\ \Eprint {http://arxiv.org/abs/0905.1151}
  {arXiv:0905.1151 [hep-th]} \BibitemShut {NoStop}%
\bibitem [{\citenamefont {Chang}\ and\ \citenamefont
  {Ma}(1969)}]{Chang:1968bh}%
  \BibitemOpen
  \bibfield  {author} {\bibinfo {author} {\bibfnamefont {S.-J.}\ \bibnamefont
  {Chang}}\ and\ \bibinfo {author} {\bibfnamefont {S.-K.}\ \bibnamefont {Ma}},\
  }\bibfield  {title} {\enquote {\bibinfo {title} {Feynman rules and quantum
  electrodynamics at infinite momentum},}\ }\href@noop {} {\bibfield  {journal}
  {\bibinfo  {journal} {Phys. Rev.}\ }\textbf {\bibinfo {volume} {180}},\
  \bibinfo {pages} {1506--1513} (\bibinfo {year} {1969})}\BibitemShut {NoStop}%
\bibitem [{\citenamefont {Yan}(1973)}]{Yan:1973qg}%
  \BibitemOpen
  \bibfield  {author} {\bibinfo {author} {\bibfnamefont {T.-M.}\ \bibnamefont
  {Yan}},\ }\bibfield  {title} {\enquote {\bibinfo {title} {Quantum field
  theories in the infinite momentum frame. 4. {S}cattering matrix of vector and
  {D}irac fields and perturbation theory},}\ }\href@noop {} {\bibfield
  {journal} {\bibinfo  {journal} {Phys. Rev.}\ }\textbf {\bibinfo {volume}
  {D7}},\ \bibinfo {pages} {1780--1800} (\bibinfo {year} {1973})}\BibitemShut
  {NoStop}%
\bibitem [{\citenamefont {Nakanishi}\ and\ \citenamefont
  {Yabuki}(1977)}]{Nakanishi:1976yx}%
  \BibitemOpen
  \bibfield  {author} {\bibinfo {author} {\bibfnamefont {N.}~\bibnamefont
  {Nakanishi}}\ and\ \bibinfo {author} {\bibfnamefont {H.}~\bibnamefont
  {Yabuki}},\ }\bibfield  {title} {\enquote {\bibinfo {title} {Null-plane
  quantization and {H}aag's theorem},}\ }\href@noop {} {\bibfield  {journal}
  {\bibinfo  {journal} {Lett. Math. Phys.}\ }\textbf {\bibinfo {volume} {1}},\
  \bibinfo {pages} {371--374} (\bibinfo {year} {1977})}\BibitemShut {NoStop}%
\bibitem [{\citenamefont {Nakanishi}\ and\ \citenamefont
  {Yamawaki}(1977)}]{Nakanishi:1976vf}%
  \BibitemOpen
  \bibfield  {author} {\bibinfo {author} {\bibfnamefont {N.}~\bibnamefont
  {Nakanishi}}\ and\ \bibinfo {author} {\bibfnamefont {K.}~\bibnamefont
  {Yamawaki}},\ }\bibfield  {title} {\enquote {\bibinfo {title} {A consistent
  formulation of the null-plane quantum field theory},}\ }\href@noop {}
  {\bibfield  {journal} {\bibinfo  {journal} {Nucl. Phys.}\ }\textbf {\bibinfo
  {volume} {B122}},\ \bibinfo {pages} {15--28} (\bibinfo {year}
  {1977})}\BibitemShut {NoStop}%
\bibitem [{\citenamefont {Haag}(1955)}]{Haag:1955ev}%
  \BibitemOpen
  \bibfield  {author} {\bibinfo {author} {\bibfnamefont {R.}~\bibnamefont
  {Haag}},\ }\bibfield  {title} {\enquote {\bibinfo {title} {On quantum field
  theories},}\ }\href@noop {} {\bibfield  {journal} {\bibinfo  {journal} {Kong.
  Dan. Vid. Sel. Mat. Fys. Med.}\ }\textbf {\bibinfo {volume} {29}},\ \bibinfo
  {pages} {1--37} (\bibinfo {year} {1955})},\ \bibinfo {note} {[Phil. Mag.
  Ser.746,376(1955)]}\BibitemShut {NoStop}%
\bibitem [{\citenamefont {Streater}\ and\ \citenamefont
  {Wightman}(1964)}]{Streater:1964}%
  \BibitemOpen
  \bibfield  {author} {\bibinfo {author} {\bibfnamefont {R.~F.}\ \bibnamefont
  {Streater}}\ and\ \bibinfo {author} {\bibfnamefont {A.~S.}\ \bibnamefont
  {Wightman}},\ }\href@noop {} {\emph {\bibinfo {title} {{PCT}, spin and
  statistics, and all that}}}\ (\bibinfo  {publisher} {W. A. Benjamin},\
  \bibinfo {year} {1964})\BibitemShut {NoStop}%
\bibitem [{\citenamefont {Heinzl}(2003)}]{Heinzl:2003jy}%
  \BibitemOpen
  \bibfield  {author} {\bibinfo {author} {\bibfnamefont {T.}~\bibnamefont
  {Heinzl}},\ }\bibfield  {title} {\enquote {\bibinfo {title} {Light-cone zero
  modes revisited},}\ }\href@noop {} {\  (\bibinfo {year} {2003})},\ \Eprint
  {http://arxiv.org/abs/hep-th/0310165} {hep-th/0310165} \BibitemShut {NoStop}%
\bibitem [{\citenamefont {Herrmann}\ and\ \citenamefont
  {Polyzou}(2015)}]{Herrmann:2015dqa}%
  \BibitemOpen
  \bibfield  {author} {\bibinfo {author} {\bibfnamefont {M.}~\bibnamefont
  {Herrmann}}\ and\ \bibinfo {author} {\bibfnamefont {W.~N.}\ \bibnamefont
  {Polyzou}},\ }\bibfield  {title} {\enquote {\bibinfo {title} {Light-front
  vacuum},}\ }\href {\doibase 10.1103/PhysRevD.91.085043} {\bibfield  {journal}
  {\bibinfo  {journal} {Phys. Rev.}\ }\textbf {\bibinfo {volume} {D91}},\
  \bibinfo {pages} {085043} (\bibinfo {year} {2015})},\ \Eprint
  {http://arxiv.org/abs/1502.01230} {arXiv:1502.01230 [hep-th]} \BibitemShut
  {NoStop}%
\bibitem [{\citenamefont {Collins}(2011)}]{Collins:2011qcdbook}%
  \BibitemOpen
  \bibfield  {author} {\bibinfo {author} {\bibfnamefont {J.~C.}\ \bibnamefont
  {Collins}},\ }\href {\doibase 10.1017/CBO9780511975592} {\emph {\bibinfo
  {title} {Foundations of Perturbative QCD}}}\ (\bibinfo  {publisher}
  {Cambridge University Press},\ \bibinfo {address} {Cambridge},\ \bibinfo
  {year} {2011})\BibitemShut {NoStop}%
\bibitem [{\citenamefont {Maskawa}\ and\ \citenamefont
  {Yamawaki}(1976)}]{Maskawa:1975ky}%
  \BibitemOpen
  \bibfield  {author} {\bibinfo {author} {\bibfnamefont {T.}~\bibnamefont
  {Maskawa}}\ and\ \bibinfo {author} {\bibfnamefont {K.}~\bibnamefont
  {Yamawaki}},\ }\bibfield  {title} {\enquote {\bibinfo {title} {The problem of
  {$P^+=0$} mode in the null-plane field theory and {D}irac's method of
  quantization},}\ }\href {\doibase 10.1143/PTP.56.270} {\bibfield  {journal}
  {\bibinfo  {journal} {Prog. Theor. Phys.}\ }\textbf {\bibinfo {volume}
  {56}},\ \bibinfo {pages} {270} (\bibinfo {year} {1976})}\BibitemShut
  {NoStop}%
\bibitem [{\citenamefont {Tsujimaru}\ and\ \citenamefont
  {Yamawaki}(1998)}]{Tsujimaru:1997jt}%
  \BibitemOpen
  \bibfield  {author} {\bibinfo {author} {\bibfnamefont {S.}~\bibnamefont
  {Tsujimaru}}\ and\ \bibinfo {author} {\bibfnamefont {K.}~\bibnamefont
  {Yamawaki}},\ }\bibfield  {title} {\enquote {\bibinfo {title} {Zero mode and
  symmetry breaking on the light front},}\ }\href {\doibase
  10.1103/PhysRevD.57.4942} {\bibfield  {journal} {\bibinfo  {journal} {Phys.
  Rev.}\ }\textbf {\bibinfo {volume} {D57}},\ \bibinfo {pages} {4942--4964}
  (\bibinfo {year} {1998})},\ \Eprint {http://arxiv.org/abs/hep-th/9704171}
  {arXiv:hep-th/9704171 [hep-th]} \BibitemShut {NoStop}%
\bibitem [{\citenamefont {Yamawaki}(1998)}]{Yamawaki:1998cy}%
  \BibitemOpen
  \bibfield  {author} {\bibinfo {author} {\bibfnamefont {K.}~\bibnamefont
  {Yamawaki}},\ }\bibfield  {title} {\enquote {\bibinfo {title} {Zero-mode
  problem on the light front},}\ }\href@noop {} {\  (\bibinfo {year} {1998})},\
  \Eprint {http://arxiv.org/abs/hep-th/9802037} {arXiv:hep-th/9802037 [hep-th]}
  \BibitemShut {NoStop}%
\bibitem [{\citenamefont {DeGennes}(1966)}]{DeGennes.1966}%
  \BibitemOpen
  \bibfield  {author} {\bibinfo {author} {\bibfnamefont {P.~G.}\ \bibnamefont
  {DeGennes}},\ }\href@noop {} {\emph {\bibinfo {title} {Superconductivity of
  Metals and Alloys}}}\ (\bibinfo  {publisher} {Addison-Wesley},\ \bibinfo
  {year} {1966})\BibitemShut {NoStop}%
\bibitem [{\citenamefont {Jain}(2007)}]{Jain2007.book}%
  \BibitemOpen
  \bibfield  {author} {\bibinfo {author} {\bibfnamefont {J.~K.}\ \bibnamefont
  {Jain}},\ }\href@noop {} {\emph {\bibinfo {title} {Composite Fermions}}}\
  (\bibinfo  {publisher} {Cambridge University Press},\ \bibinfo {address}
  {Cambridge},\ \bibinfo {year} {2007})\BibitemShut {NoStop}%
\bibitem [{\citenamefont {Alberg}\ and\ \citenamefont
  {Miller}(2017)}]{Alberg:2017ijg}%
  \BibitemOpen
  \bibfield  {author} {\bibinfo {author} {\bibfnamefont {M.}~\bibnamefont
  {Alberg}}\ and\ \bibinfo {author} {\bibfnamefont {G.~A.}\ \bibnamefont
  {Miller}},\ }\bibfield  {title} {\enquote {\bibinfo {title} {Chiral light
  front perturbation theory and the flavor dependence of the light-quark
  nucleon sea},}\ }\href@noop {} {\  (\bibinfo {year} {2017})},\ \Eprint
  {http://arxiv.org/abs/1712.05814} {arXiv:1712.05814 [nucl-th]} \BibitemShut
  {NoStop}%
\bibitem [{\citenamefont {Hiller}(2016)}]{Hiller:2016itl}%
  \BibitemOpen
  \bibfield  {author} {\bibinfo {author} {\bibfnamefont {J.~R.}\ \bibnamefont
  {Hiller}},\ }\bibfield  {title} {\enquote {\bibinfo {title} {Nonperturbative
  light-front {H}amiltonian methods},}\ }\href {\doibase
  10.1016/j.ppnp.2016.06.002} {\bibfield  {journal} {\bibinfo  {journal} {Prog.
  Part. Nucl. Phys.}\ }\textbf {\bibinfo {volume} {90}},\ \bibinfo {pages}
  {75--124} (\bibinfo {year} {2016})},\ \Eprint
  {http://arxiv.org/abs/1606.08348} {arXiv:1606.08348 [hep-ph]} \BibitemShut
  {NoStop}%
\bibitem [{\citenamefont {Vary}\ \emph {et~al.}(2017)\citenamefont {Vary} \emph
  {et~al.}}]{Vary:2016ccz}%
  \BibitemOpen
  \bibfield  {author} {\bibinfo {author} {\bibfnamefont {J.~P.}\ \bibnamefont
  {Vary}} \emph {et~al.},\ }\bibfield  {title} {\enquote {\bibinfo {title}
  {Trends and progress in nuclear and hadron physics: a straight or winding
  road},}\ }\bibfield  {booktitle} {\emph {\bibinfo {booktitle} {{Proceedings,
  Theory and Experiment for Hadrons on the Light-Front (Light Cone 2016):
  Lisbon, Portugal, September 5-8, 2016}}},\ }\href {\doibase
  10.1007/s00601-016-1210-1} {\bibfield  {journal} {\bibinfo  {journal} {Few
  Body Syst.}\ }\textbf {\bibinfo {volume} {58}},\ \bibinfo {pages} {56}
  (\bibinfo {year} {2017})},\ \Eprint {http://arxiv.org/abs/1612.03963}
  {arXiv:1612.03963 [nucl-th]} \BibitemShut {NoStop}%
\bibitem [{\citenamefont {Vary}\ \emph {et~al.}(2016)\citenamefont {Vary},
  \citenamefont {Adhikari}, \citenamefont {Chen}, \citenamefont {Li},
  \citenamefont {Maris},\ and\ \citenamefont {Zhao}}]{Vary:2016emi}%
  \BibitemOpen
  \bibfield  {author} {\bibinfo {author} {\bibfnamefont {J.~P.}\ \bibnamefont
  {Vary}}, \bibinfo {author} {\bibfnamefont {L.}~\bibnamefont {Adhikari}},
  \bibinfo {author} {\bibfnamefont {G.}~\bibnamefont {Chen}}, \bibinfo {author}
  {\bibfnamefont {Y.}~\bibnamefont {Li}}, \bibinfo {author} {\bibfnamefont
  {P.}~\bibnamefont {Maris}}, \ and\ \bibinfo {author} {\bibfnamefont
  {X.}~\bibnamefont {Zhao}},\ }\bibfield  {title} {\enquote {\bibinfo {title}
  {{Basis Light-Front Quantization: Recent Progress and Future Prospects}},}\
  }\bibfield  {booktitle} {\emph {\bibinfo {booktitle} {{Proceedings, Theory
  and Experiment for Hadrons on the Light-Front (Light Cone 2015): Frascati ,
  Italy, September 21-25, 2015}}},\ }\href {\doibase 10.1007/s00601-016-1117-x}
  {\bibfield  {journal} {\bibinfo  {journal} {Few Body Syst.}\ }\textbf
  {\bibinfo {volume} {57}},\ \bibinfo {pages} {695--702} (\bibinfo {year}
  {2016})}\BibitemShut {NoStop}%
\bibitem [{\citenamefont {Brown}(1992)}]{Brown:1992db}%
  \BibitemOpen
  \bibfield  {author} {\bibinfo {author} {\bibfnamefont {L.~S.}\ \bibnamefont
  {Brown}},\ }\href@noop {} {\emph {\bibinfo {title} {Quantum Field Theory}}}\
  (\bibinfo  {publisher} {Cambridge University Press},\ \bibinfo {address}
  {Cambridge},\ \bibinfo {year} {1992})\BibitemShut {NoStop}%
\bibitem [{\citenamefont {Schlieder}\ and\ \citenamefont
  {Seiler}(1972)}]{Schlieder:1972qr}%
  \BibitemOpen
  \bibfield  {author} {\bibinfo {author} {\bibfnamefont {S.}~\bibnamefont
  {Schlieder}}\ and\ \bibinfo {author} {\bibfnamefont {E.}~\bibnamefont
  {Seiler}},\ }\bibfield  {title} {\enquote {\bibinfo {title} {Remarks on the
  null plane development of a relativistic quantum field theory},}\ }\href
  {\doibase 10.1007/BF01877587} {\bibfield  {journal} {\bibinfo  {journal}
  {Commun. Math. Phys.}\ }\textbf {\bibinfo {volume} {25}},\ \bibinfo {pages}
  {62--72} (\bibinfo {year} {1972})}\BibitemShut {NoStop}%
\bibitem [{\citenamefont {Kogut}\ and\ \citenamefont
  {Soper}(1970)}]{Kogut:1969xa}%
  \BibitemOpen
  \bibfield  {author} {\bibinfo {author} {\bibfnamefont {J.~B.}\ \bibnamefont
  {Kogut}}\ and\ \bibinfo {author} {\bibfnamefont {D.~E.}\ \bibnamefont
  {Soper}},\ }\bibfield  {title} {\enquote {\bibinfo {title} {Quantum
  electrodynamics in the infinite momentum frame},}\ }\href@noop {} {\bibfield
  {journal} {\bibinfo  {journal} {Phys. Rev.}\ }\textbf {\bibinfo {volume}
  {D1}},\ \bibinfo {pages} {2901--2913} (\bibinfo {year} {1970})}\BibitemShut
  {NoStop}%
\bibitem [{\citenamefont {Ma}\ and\ \citenamefont {Wang}(2006)}]{Ma:2006dp}%
  \BibitemOpen
  \bibfield  {author} {\bibinfo {author} {\bibfnamefont {J.~P.}\ \bibnamefont
  {Ma}}\ and\ \bibinfo {author} {\bibfnamefont {Q.}~\bibnamefont {Wang}},\
  }\bibfield  {title} {\enquote {\bibinfo {title} {On transverse-momentum
  dependent light-cone wave functions of light mesons},}\ }\href {\doibase
  10.1016/j.physletb.2006.09.029} {\bibfield  {journal} {\bibinfo  {journal}
  {Phys. Lett.}\ }\textbf {\bibinfo {volume} {B642}},\ \bibinfo {pages}
  {232--237} (\bibinfo {year} {2006})},\ \Eprint
  {http://arxiv.org/abs/hep-ph/0605075} {arXiv:hep-ph/0605075 [hep-ph]}
  \BibitemShut {NoStop}%
\bibitem [{\citenamefont {Brodsky}\ \emph {et~al.}(2001)\citenamefont
  {Brodsky}, \citenamefont {Hwang}, \citenamefont {Ma},\ and\ \citenamefont
  {Schmidt}}]{Brodsky:2000ii}%
  \BibitemOpen
  \bibfield  {author} {\bibinfo {author} {\bibfnamefont {S.~J.}\ \bibnamefont
  {Brodsky}}, \bibinfo {author} {\bibfnamefont {D.-S.}\ \bibnamefont {Hwang}},
  \bibinfo {author} {\bibfnamefont {B.-Q.}\ \bibnamefont {Ma}}, \ and\ \bibinfo
  {author} {\bibfnamefont {I.}~\bibnamefont {Schmidt}},\ }\bibfield  {title}
  {\enquote {\bibinfo {title} {Light-cone representation of the spin and
  orbital angular momentum of relativistic composite systems},}\ }\href@noop {}
  {\bibfield  {journal} {\bibinfo  {journal} {Nucl. Phys.}\ }\textbf {\bibinfo
  {volume} {B593}},\ \bibinfo {pages} {311--335} (\bibinfo {year} {2001})},\
  \Eprint {http://arxiv.org/abs/hep-th/0003082} {hep-th/0003082} \BibitemShut
  {NoStop}%
\bibitem [{\citenamefont {Collins}\ and\ \citenamefont
  {Soper}(1981)}]{Collins:1981uk}%
  \BibitemOpen
  \bibfield  {author} {\bibinfo {author} {\bibfnamefont {J.~C.}\ \bibnamefont
  {Collins}}\ and\ \bibinfo {author} {\bibfnamefont {D.~E.}\ \bibnamefont
  {Soper}},\ }\bibfield  {title} {\enquote {\bibinfo {title} {Back-to-back jets
  in {QCD}},}\ }\href {\doibase 10.1016/0550-3213(81)90339-4} {\bibfield
  {journal} {\bibinfo  {journal} {Nucl. Phys.}\ }\textbf {\bibinfo {volume}
  {B193}},\ \bibinfo {pages} {381--443} (\bibinfo {year} {1981})},\ \bibinfo
  {note} {erratum: \BIBvol{B213}, 545 (1983)}\BibitemShut {NoStop}%
\bibitem [{\citenamefont {Collins}\ and\ \citenamefont
  {Soper}(1982)}]{Collins:1981uw}%
  \BibitemOpen
  \bibfield  {author} {\bibinfo {author} {\bibfnamefont {J.~C.}\ \bibnamefont
  {Collins}}\ and\ \bibinfo {author} {\bibfnamefont {D.~E.}\ \bibnamefont
  {Soper}},\ }\bibfield  {title} {\enquote {\bibinfo {title} {Parton
  distribution and decay functions},}\ }\href {\doibase
  10.1016/0550-3213(82)90021-9} {\bibfield  {journal} {\bibinfo  {journal}
  {Nucl. Phys.}\ }\textbf {\bibinfo {volume} {B194}},\ \bibinfo {pages}
  {445--492} (\bibinfo {year} {1982})}\BibitemShut {NoStop}%
\bibitem [{\citenamefont {Soper}(1979)}]{Soper:1979fq}%
  \BibitemOpen
  \bibfield  {author} {\bibinfo {author} {\bibfnamefont {D.~E.}\ \bibnamefont
  {Soper}},\ }\bibfield  {title} {\enquote {\bibinfo {title} {Partons and their
  transverse momenta in {QCD}},}\ }\href@noop {} {\bibfield  {journal}
  {\bibinfo  {journal} {Phys. Rev. Lett.}\ }\textbf {\bibinfo {volume} {43}},\
  \bibinfo {pages} {1847--1851} (\bibinfo {year} {1979})}\BibitemShut {NoStop}%
\bibitem [{\citenamefont {Becher}\ and\ \citenamefont
  {Neubert}(2011)}]{Becher:2010tm}%
  \BibitemOpen
  \bibfield  {author} {\bibinfo {author} {\bibfnamefont {T.}~\bibnamefont
  {Becher}}\ and\ \bibinfo {author} {\bibfnamefont {M.}~\bibnamefont
  {Neubert}},\ }\bibfield  {title} {\enquote {\bibinfo {title} {{D}rell-{Y}an
  production at small {$q_T$}, transverse parton distributions and the
  collinear anomaly},}\ }\href {\doibase 10.1140/epjc/s10052-011-1665-7}
  {\bibfield  {journal} {\bibinfo  {journal} {Eur. Phys. J.}\ }\textbf
  {\bibinfo {volume} {C71}},\ \bibinfo {pages} {1665} (\bibinfo {year}
  {2011})},\ \Eprint {http://arxiv.org/abs/1007.4005} {arXiv:1007.4005
  [hep-ph]} \BibitemShut {NoStop}%
\bibitem [{\citenamefont {Echevarr\'{\i}a}\ \emph {et~al.}(2012)\citenamefont
  {Echevarr\'{\i}a}, \citenamefont {Idilbi},\ and\ \citenamefont
  {Scimemi}}]{GarciaEchevarria:2011rb}%
  \BibitemOpen
  \bibfield  {author} {\bibinfo {author} {\bibfnamefont {M.~G.}\ \bibnamefont
  {Echevarr\'{\i}a}}, \bibinfo {author} {\bibfnamefont {A.}~\bibnamefont
  {Idilbi}}, \ and\ \bibinfo {author} {\bibfnamefont {I.}~\bibnamefont
  {Scimemi}},\ }\bibfield  {title} {\enquote {\bibinfo {title} {Factorization
  theorem for {D}rell-{Y}an at low {$q_T$} and transverse momentum
  distributions on-the-light-cone},}\ }\href {\doibase 10.1007/JHEP07(2012)002}
  {\bibfield  {journal} {\bibinfo  {journal} {JHEP}\ }\textbf {\bibinfo
  {volume} {1207}},\ \bibinfo {pages} {002} (\bibinfo {year} {2012})},\ \Eprint
  {http://arxiv.org/abs/1111.4996} {arXiv:1111.4996 [hep-ph]} \BibitemShut
  {NoStop}%
\bibitem [{\citenamefont {Collins}\ and\ \citenamefont
  {Rogers}(2013)}]{Collins:2012uy}%
  \BibitemOpen
  \bibfield  {author} {\bibinfo {author} {\bibfnamefont {J.~C.}\ \bibnamefont
  {Collins}}\ and\ \bibinfo {author} {\bibfnamefont {T.~C.}\ \bibnamefont
  {Rogers}},\ }\bibfield  {title} {\enquote {\bibinfo {title} {Equality of two
  definitions for transverse momentum dependent parton distribution
  functions},}\ }\href {\doibase 10.1103/PhysRevD.87.034018} {\bibfield
  {journal} {\bibinfo  {journal} {Phys. Rev.}\ }\textbf {\bibinfo {volume}
  {D87}},\ \bibinfo {pages} {034018} (\bibinfo {year} {2013})},\ \Eprint
  {http://arxiv.org/abs/1210.2100} {arXiv:1210.2100 [hep-ph]} \BibitemShut
  {NoStop}%
\bibitem [{\citenamefont {Li}\ and\ \citenamefont {Wang}(2015)}]{Li:2014xda}%
  \BibitemOpen
  \bibfield  {author} {\bibinfo {author} {\bibfnamefont {H.-n.}\ \bibnamefont
  {Li}}\ and\ \bibinfo {author} {\bibfnamefont {Y.-M.}\ \bibnamefont {Wang}},\
  }\bibfield  {title} {\enquote {\bibinfo {title} {Non-dipolar {W}ilson links
  for transverse-momentum-dependent wave functions},}\ }\href {\doibase
  10.1007/JHEP06(2015)013} {\bibfield  {journal} {\bibinfo  {journal} {JHEP}\
  }\textbf {\bibinfo {volume} {06}},\ \bibinfo {pages} {013} (\bibinfo {year}
  {2015})},\ \Eprint {http://arxiv.org/abs/1410.7274} {arXiv:1410.7274
  [hep-ph]} \BibitemShut {NoStop}%
\bibitem [{\citenamefont {Radyushkin}(1977)}]{Radyushkin:1977gp}%
  \BibitemOpen
  \bibfield  {author} {\bibinfo {author} {\bibfnamefont {A.~V.}\ \bibnamefont
  {Radyushkin}},\ }\bibfield  {title} {\enquote {\bibinfo {title} {Deep elastic
  processes of composite particles in field theory and asymptotic freedom},}\
  }\href@noop {} {\  (\bibinfo {year} {1977})},\ \Eprint
  {http://arxiv.org/abs/hep-ph/0410276} {arXiv:hep-ph/0410276 [hep-ph]}
  \BibitemShut {NoStop}%
\bibitem [{\citenamefont {Radyushkin}(2009)}]{Radyushkin:2009zg}%
  \BibitemOpen
  \bibfield  {author} {\bibinfo {author} {\bibfnamefont {A.~V.}\ \bibnamefont
  {Radyushkin}},\ }\bibfield  {title} {\enquote {\bibinfo {title} {Shape of
  pion distribution amplitude},}\ }\href {\doibase 10.1103/PhysRevD.80.094009}
  {\bibfield  {journal} {\bibinfo  {journal} {Phys. Rev.}\ }\textbf {\bibinfo
  {volume} {D80}},\ \bibinfo {pages} {094009} (\bibinfo {year} {2009})},\
  \Eprint {http://arxiv.org/abs/0906.0323} {arXiv:0906.0323 [hep-ph]}
  \BibitemShut {NoStop}%
\bibitem [{\citenamefont {Lepage}\ and\ \citenamefont
  {Brodsky}(1979)}]{Lepage:1979zb}%
  \BibitemOpen
  \bibfield  {author} {\bibinfo {author} {\bibfnamefont {G.~P.}\ \bibnamefont
  {Lepage}}\ and\ \bibinfo {author} {\bibfnamefont {S.~J.}\ \bibnamefont
  {Brodsky}},\ }\bibfield  {title} {\enquote {\bibinfo {title} {Exclusive
  processes in quantum chromodynamics: Evolution equations for hadronic wave
  functions and the form-factors of mesons},}\ }\href {\doibase
  10.1016/0370-2693(79)90554-9} {\bibfield  {journal} {\bibinfo  {journal}
  {Phys. Lett.}\ }\textbf {\bibinfo {volume} {87B}},\ \bibinfo {pages}
  {359--365} (\bibinfo {year} {1979})}\BibitemShut {NoStop}%
\bibitem [{\citenamefont {Lepage}\ and\ \citenamefont
  {Brodsky}(1980)}]{Lepage:1980fj}%
  \BibitemOpen
  \bibfield  {author} {\bibinfo {author} {\bibfnamefont {G.~P.}\ \bibnamefont
  {Lepage}}\ and\ \bibinfo {author} {\bibfnamefont {S.~J.}\ \bibnamefont
  {Brodsky}},\ }\bibfield  {title} {\enquote {\bibinfo {title} {Exclusive
  processes in perturbative quantum chromodynamics},}\ }\href@noop {}
  {\bibfield  {journal} {\bibinfo  {journal} {Phys. Rev.}\ }\textbf {\bibinfo
  {volume} {D22}},\ \bibinfo {pages} {2157--2198} (\bibinfo {year}
  {1980})}\BibitemShut {NoStop}%
\end{thebibliography}%


\end{document}